\documentstyle[12pt]{article}    
\begin{document}           
\rm
\baselineskip=0.33333in
\begin{quote} \raggedleft TAUP 2163-94
\end{quote}
\vglue 0.5in
\begin{center}{\bf Strong and Electromagnetic Interactions }
\end{center}
\begin{center}E. Comay
\end{center}

\begin{center}
School of Physics and Astronomy \\
Raymond and Beverly Sackler Faculty \\
of Exact Sciences \\
Tel Aviv University \\
Tel Aviv 69978 \\
Israel
\end{center}
\vglue 0.5in
\noindent
\vglue 0.5in
\noindent
\vglue 0.5in
\noindent
Abstract:

   Elements of a new model of strong interactions are described.
The model is based on an extension of electrodynamics that is
derived from a regular Lagrangian density. Here the electric
and magnetic fields of Maxwell equations play a symmetric role.
It is shown that results are in accordance with
general properties of nuclear and nucleon systems.

\newpage
\noindent
{\bf 1. Introduction}

    It is well known that a physical theory describes, interprets and
predicts the behavior of an objective physical reality. One result of
this meaning of a theory is that the existence of one kind of theory
does not exclude another theory that provides a
different interpretation
of the objective world. Therefore, the standard tests of the goodness
of a theory are its fit to experimental data and its logical
self-consistency. On the other hand, the
(relative) success of one theory {\em cannot}
be used as an argument against a different theory.

    The history of science in general and of physics in particular
provides many examples of different theories interpreting the same
kind of physical objects. The long controversy concerning the
structure of light is such an example. Assuming that a wave structure
of an entity contradicts its corpuscular properties, scientists argued
for more than two centuries in favor of one of these theories
of light and against the
other one. Thus, following Newton, most scientists have
supported the corpuscular theory of light for more than one hundred
years. Later, in the 19th century, the wave theory of light prevailed.
It is only after the works of Plank on black body radiation, of
Einstein on the photoelectric effect and, finally, the establishment
of quantum mechanics, that people realized that wave phenomena
do not contradict the corpuscular structure of particles in
general and of light in particular. This is an
example where two apparently contradictory theories merge. Obviously,
the final relation between two alternative theories may be different.

    A proof of the goodness of a new theory
of strong interactions is a tremendous
assignment that involves detailed calculations
of many kinds of processes whose results
should fit experimental data. On the other hand, in
any field of physics, a new incorrect
theory can generally be more easily refuted by means of
qualitative arguments. Thus, for example, assume that one looks
at the interaction between a positive charge and a negative
one. Finding the attractive force between these charges and being
inspired by gravitation, he comes out with a "theory" which says that
"all charges attract each other". Obviously, this "theory" can be
refuted by an examination of the interaction between two charges of
the same sign. In this case one does not need to use quantitative
properties
of the repulsive force in order to reject the
electromagnetic "theory" mentioned
above.

    As a first attempt in a new direction, the present work discusses
several qualitative properties of nuclei and nucleons.
It shows that the new approach described here fits well
known basic properties of these systems.
These results encourage a further investigation
in the direction presented here.

  The structure of this work is as follows.
Similarities between strong and electromagnetic
interactions are discussed in the second section. The third section
presents some general properties of a new
theory of charge-monopole systems.
The $qq$ force is discussed
in the fourth section. Results of scattering experiments
of electrons and photons on nucleons are discussed in the
fifth section. The sixth section reviews the
correspondence between nuclear forces and molecular ones.
The seventh section is devoted to the EMC effect
where nucleons swell inside nuclei.
The nuclear tensor force is discussed in the eighth section.
In the ninth section it is shown that exotic states of
strongly bound hadrons
should not exist. Some concluding remarks are
the contents of the last section.

\newpage
\noindent
{\bf 2. Strong and Electromagnetic Interactions}
\vglue 0.333333in

    Let us examine the following striking evidence concerning strong
and electromagnetic interactions. Among the four established kinds of
interactions, gravitation depends just on the mass of
particles whereas
the strong, electromagnetic and weak interactions
are sensitive to specific
properties of the interacting particles. Table 1 shows the validity
of two conservation laws under the latter kind of interactions.

   Flavor conservation indicates a qualitative
physical property of interactions. It
turns out that strong as well as electromagnetic interactions do not
alter the flavor of a closed system of interacting particles. Thus,
these interactions induce transitions of existing constituents
(namely, quarks and massive leptons)
between different configurations of
energy states. It should be noted that
strong and electromagnetic
interactions can induce pair production of massive leptons and quarks.
In the case of quarks, these processes take
the form of meson production or of baryon pair production.
However, as shown by Dirac, the electron (quark) of the
pair is considered as a particle ejected from the sea of
negative energy states whereas the positron (antiquark) of
the pair represents the hole left in the sea by
the created electron (quark).
It can be concluded that strong and electromagnetic interactions
can alter the configuration of a system of elementary particles while
the overall flavor is conserved.

    Consider for example
the weak decay of the neutron $n\rightarrow p+e+\bar {\nu }_e$.
In this process a $d$ quark is destroyed and a $u$ quark is
created. Analogous consequences can be seen in other
weak interactions. These flavor nonconservation transitions are
clearly outside the scope of strong and electromagnetic interactions,
because the latter "see" quarks, electrons etc. as elementary
indestructible entities.

    Parity conservation is another common property of strong and
electromagnetic interactions. This property indicates
mathematical symmetries of the Hamiltonians of these interactions.
On the other hand, as is well known, weak interactions do not
conserve parity.

    The foregoing brief discussions of flavor and parity
conservation in strong and electromagnetic
interactions can be concluded as follows: experiment teaches
us that strong and electromagnetic interactions
have in common some fundamental properties, namely,
flavor and parity conservation.
Moreover, there are just two kinds of basic interactions
that have these properties. This conclusion
motivates the search for a common (or similar) basis
for the strong and electromagnetic interactions. This search is the
main objective of the present work.

\vglue 0.666666in
\noindent
{\bf 3. Magnetic Monopoles}
\vglue 0.333333in

The idea that monopoles are constituents of hadrons is not new. It has
already been suggested by Dirac[1] even before quarks have
been discovered. Schwinger[2] has proposed a model of hadrons where
quarks are dyons (namely, particles having both electric and magnetic
charges) with a magnetic charge $\pm g_0$ or $\pm 2g_0$ and
\begin{equation}
g_0^2=1/4e^2\simeq 34.
\label{eq:K1}
\end{equation}
Units where $\hbar=c=1$ are used.
Barut[3] discussed the applicability of the group $O(4,2)$ for a
system of dyons. Sawada[4] claims in a series of works that a strong
force which is analogous to the van der Waals force between
molecules, is found in the hadron-hadron scattering amplitude. He
interprets this force as an indication that quarks are dyons and
that strong interactions are associated with magnetic monopoles.

    The present work relies on a
new classical theory of charges and monopoles[5].
This theory is based on the following approach.
Consider the fact that so far there is no experimental evidence
for the existence of monopoles in general
and for their equations of motion
in particular. This experimental situation means that one is not
bound to start building the theory from a specific
form of monopoles equations of motion.
On the basis of this state of affairs,
the new theory postulates that the equations of
motion of a charge-monopole system
should be derivable from a regular Lagrangian density. In this
way one obtains a charge-monopole
theory that conserves the structure of ordinary classical
electrodynamics of charges and fields[6]. Moreover, the standard way
of constructing the corresponding quantum mechanical theory can
be readily used[7,8].

    This approach utilizes few additional
self-evident postulates and arrives at
the following form of the electromagnetic part of the Lagrangian
density[5]
\begin{eqnarray}
L & = & -\frac {1}{16\pi }F_{(e,w)}^{\mu \nu }F_{(e,w)\mu \nu } -
 J_{(e)}^\mu A_{(e,w)\mu }
\nonumber \\
  &   & -\frac {1}{16\pi }F_{(m,w)}^{\mu \nu }F_{(m,w)\mu \nu } -
 J_{(m)}^\mu A_{(m,w)\mu } +
\frac {1}{16\pi }F_{(w)}^{\mu \nu }F_{(w)\mu \nu }
\label{eq:L1}
\end{eqnarray}
Here Greek indices range from 0 to 3, the metric
$g_{\mu \nu}$ is diagonal
and its entries are (1,-1,-1,-1). The subscripts $e, m$ and $w$
denote quantities associated with charges, monopoles and free
electromagnetic waves (namely, real photons), respectively.

    Using a standard mathematical method[6,5], one obtains from this
Lagrangian density the appropriate form of the Lorentz force
exerted on charges and monopoles:
\begin{equation}
Ma_{(e)}^\mu = qF_{(e,w)}^{\mu \nu }v_{(e)\nu}
\label{eq:L2}
\end{equation}
\begin{equation}
Ma_{(m)}^\mu = qF_{(m,w)}^{\mu \nu }v_{(m)\nu}
\label{eq:L3}
\end{equation}

Results $(\!\!~\ref{eq:L2})$ and
$(\!\!~\ref{eq:L3})$ of this theory can be put as follows:
\begin{itemize}
\item[{(I)}]
Charges do not interact with fields of monopoles;
monopoles do not interact with fields of charges; charges and
monopoles interact indirectly through fields of real photons.
\end{itemize}
Henceforth, these results are denoted by (I).
Properties (I) are derived theoretically
within the framework of classical
electrodynamics. They rely upon a successful approach to
the foundations of physics, namely the variational principle.

The possibility that the foregoing results are
related to the real world is discussed in this work. Fortunately,
(I) has an immediate correspondence to
experimental data which can be put in the following statement:
\begin{itemize}
\item[{(II)}]
Electrons, as well as other charged leptons, do not participate in
strong interactions.
\end{itemize}
This well established property of Nature
corresponds to (I) if one regards electrons
as pure charges, quarks as dyons and strong interactions
as interactions between magnetic monopoles. This
clear correspondence between experiment and theory
provides an encouragement for a further investigation
of the possibility that quarks are dyons.

   Another result of reference 5 is used here. Unlike other monopole
theories, the new theory does not impose any restriction on the
strength of the elementary magnetic charge $g_0$
and leaves it as a free parameter. This conclusion is related
to the following property of hadrons. The examination
of a $q\bar q$ bound states where the two quarks are of the
$u,d$ flavors, shows pions of 135 Mev and other bound states of
such mesons which are many times more
massive. On the other hand, all states of
the positronium, namely the $e\bar e$ bound states,
vary within an energy range which is about $10^{-5}$
of the system's mass. This evidence indicates
that the elementary magnetic charge should be much stronger than its
electric counterpart. However, the extremely high value
$g_0^2\simeq 34$ of $(\!\!~\ref{eq:K1})$
is not imposed on any specific calculation
associated with this work.

    Another aspect of the fit of a charge-monopole system, as
a theoretical interpretation of Nature, is as follows.
The behavior of a pure monopole system is obtained from that of
a pure electric one by the transformations
\begin{equation}
\mbox{\boldmath $E$}\rightarrow \mbox{\boldmath $B$}; \;\;\;
\mbox{\boldmath $B$}\rightarrow -\mbox{\boldmath $E$}; \;\;\;
e\rightarrow g.
\label{eq:DUALITY}
\end{equation}
These relations show that there are no more than
two kinds of related charges called electric charge and magnetic
monopole. This
point is in accordance with Nature which shows only two kinds of
fundamental interactions that conserve flavor and parity, namely,
electromagnetic and strong interactions. This point is explained
here probably for the first time.

The previous discussion can
be put in a different way. The theory presented here predicts
the existence of two related kinds of fundamental interactions
that conserve
flavor and parity. This prediction is consistent with Nature.
It could have been refuted if, for example, experiment would
show {\em three} different kinds of fundamental interactions that
share these properties. Thus, the success of the ideas
described here in this simple but essential aspect, motivates
a further research in this direction.

\vglue 0.666666in
\noindent
{\bf 4. The qq Force}
\vglue 0.333333in
    Quantum numbers of mesons are consistent with those
of bound states of a $q\bar q$ pair.
Hence, the $q\bar q$
force should be attractive. This property corresponds to the
attraction between electric particles of opposite signs.
These states are analogous to those of the positronium
where an electron and a positron are bound together.

   As pointed out above, electrodynamics of monopoles is
completely analogous to that of charges. Hence, one expects
that the $qq$ force should be repulsive. As is well known,
quantum numbers of baryons are derived if baryons are regarded as
systems of three quarks. An explanation how these
quarks form a bound state in spite of the repulsive $qq$ force, can
be borrowed from the atomic system: a baryon has a core whose
magnetic charge is $+3g_0$, to which three quarks are attracted.
The magnetic charge of each quark is $-g_0$ and the overall magnetic
charge of the system vanishes. (Hereafter $-g_0$ denotes the unit
magnetic charge ascribed to a quark and not the huge value
$(\!\!~\ref{eq:K1})$.) Quantum states of baryons are
known to be characterized by these three quarks. The last statement
is analogous to the one saying that an atomic quantum state
is characterized by its electronic part, namely a statement which
is correct if effects of hyperfine interactions
are ignored.

    The idea that baryons have a core is not new. It has already
been used in early baryonic models[9]. Gluons of QCD, which
bear strong charge, provide another example of a charged baryonic core
used in other models.

    An indication that baryons have a core can be found in data
on quarks' portion of the nucleon's momentum. Experiments
show that this
portion is about one half of the entire momentum of the
nucleon[10]. Hence, nucleons are made of quarks and of other
constituents. The latter are called here the baryonic core.
It is shown below that other properties of nuclear
and nucleon interactions are compatible with the
theory used here and {\em are explained without any further
assumption}.

    An additional experimental evidence supporting the postulate
of baryonic structure described above is the momentum
distribution of antiquarks in nucleons. This distribution
is inferred from the width of their structure
function as plotted in terms of Bjorken's $x$[10]. If one
compares this width to that of quarks then he finds
that antiquarks have
a smaller Fermi motion. This outcome means that antiquarks are
spread in a larger spatial volume than quarks do. This property
is explained in a simple manner if one assumes that nucleons have
a core whose strong charge is $3g_0$. Under the assumption that
strong charge is analogous to electromagnetic one, it is expected
that quarks,
whose charge equals $-g_0$, are attracted to the
core whereas antiquarks are repelled by it. Near the core, quarks
do not completely screen the core's field and antiquarks are
repelled from that region to outer portions of the nucleonic
volume. Hence, due to the uncertainty principle,
one infers that a smaller
Fermi motion should be found in experimental data of antiquarks
in nucleons.

    It can be concluded that hadronic features discussed
here do not contradict the fundamental electrodynamic property
where charges having the same sign repel each other. This
property is manifested here in a new form where quarks are
considered as magnetic monopoles.

\vglue 0.666666in
\noindent
{\bf 5. Scattering of Electrons and Photons on Nucleons}
\vglue 0.333333in

    Electrons and photons are electromagnetic entities. High
energy scattering processes of these particles on nucleons
reveal properties of nucleon constituents. In this section,
results of deep inelastic electron scattering on protons
are compaired to those of neutrons. The same is done
for hard $\gamma $ photons.
In the case of electrons, one finds that the
total cross section of protons, $\sigma _{p}$, is greater
than that of neutrons $\sigma _{n}$[11]. Indeed, the following
neutron/proton structure function
ratio is described well by the decreasing line
\begin{equation}
F_2^{en}/F_2^{ep} = 0.96-0.75x
\label{eq:NP}
\end{equation}
(see fig. 7 in the article of [11] and the book
mentioned there).
Hence, since the mean value of $x$ is positive,
one finds that the cross section of high energy
electrons scattered off protons is greater than that of neutrons.
On the other hand, the total cross section of hard photon
scattering is the same for protons and neutrons[12].

    These results indicate that high energy electrons
"see" different electric charges inside the two
kinds of nucleons. On the other hand, photons do not
discriminate between protons and neutrons. This outcome is
consistent with the Lagrangian $(\!\!~\ref{eq:L1})$ and
the equations of motion $(\!\!~\ref{eq:L2})$ and
$(\!\!~\ref{eq:L3})$. Indeed, the electron, which is
a pure electric charge "sees" only the electric charges of
the quarks in each kind of nucleons. Hence, because
of the variation in nucleonic electric charge, electrons interact
differently with constituents of protons and neutrons.
On the other hand, photons "see" both electric charges
and magnetic monopoles. Due to the strong magnetic charge
ascribed to quarks, their electric charge can be ignored
in an analysis of their interaction with hard photons. Hence,
the proton and the neutron look alike in experiments
with hard photons.

    The foregoing discussion explains the different behavior
of high energy electrons and photons scattered on protons and
neutrons.

\newpage
\noindent
{\bf 6. The Correspondence Between Nuclear and Molecular Forces}
\vglue 0.333333in

    There are many similarities between nuclear forces and molecular
ones. Molecular forces are relatively weak and are associated with
interactions between electrically neutral molecules. In the following
lines it is shown that analogous forces exist between nucleons in a
nucleus.

    According to the interpretation
presented here, nucleons consist of magnetically
charged particles in a way where the overall magnetic charge vanishes.
Similarly, molecules are systems whose constituents are electrically
charged particles having a null electric charge. This is
the origin of the similarity between nuclear and molecular forces.
Before turning to effects that demonstrate this similarity,
let us point
out several issues which show that the analogy between the two
kinds of systems bound by these residual forces is
not complete.

    Molecules are neutral with respect to both electric charges and
magnetic monopoles. Nucleons, on the other hand, are neutral only with
respect to the stronger unit of elementary charge, namely the
magnetic monopole. The net electric charge of protons is the
origin of nuclear Coulomb interaction which increases with the number
of protons in nuclei. This force is the reason for the instability
of high $Z$ nuclei. No analogous property is found in molecular
forces where a liquid can take a macroscopic size.

    There is only one kind of electrons in stable atoms. On the
other hand, there are two kinds of valence quarks in nucleons,
namely the $u$ and $d$ quarks. This is the
basis of isospin symmetry found in nuclear interactions. There
is no counterpart to this symmetry in molecular interactions.

    Remembering these points, let us turn to analogous features
of the two forces discussed in this section.

    The two forces are of a residual nature. The
nuclear force is much weaker
than the corresponding strong force between dyons and
molecular forces are similarly related to Coulomb interactions between
electrons. Indeed, the binding energy of a typical nucleus is about 8
Mev per nucleon whereas excited states of nucleons and of mesons
are measured by hundreds of Mev. A similar relation holds in
molecules. The binding energy of a molecule in a liquid is generally
a fraction of ev. On the other hand, an excited state of electrons
is of the order of 10 ev.

    Both nuclear forces and molecular ones are characterized by a
hard core, outside of which there exists an attractive force which
falls off much faster than a Coulomb force. This is the reason for
the constant density of nuclei and of liquids, a property
which explains
the success of nuclear liquid drop models[13,14].

\vglue 0.666666in
\noindent
{\bf 7. The EMC Effect}
\vglue 0.333333in

    The size of the nucleonic volume can be deduced from the
nucleonic quarks' momentum. It is found that the larger the
number of nucleons $A$ in a nucleus the larger is the mean
self volume of nucleons of this nucleus. In other words, as
the nucleus becomes heavier its nucleons swell. This property
is compatible with the EMC effect[15]. A support for the
swelling effect
is found in a measurement of the cross section of $K^+$
scattering on nuclei[16]. On the other hand, the success
of the nuclear liquid drop models is an indication that nuclear
density is practically constant[17].

    The swelling of the mean volume of a nucleon with the
increase of the nuclear number $A$ is compatible with
screening properties of electrodynamics. Consider a nucleon
$N_i$ in a nucleus. A part of
the wave function of quarks of neighboring
nucleons penetrates into the volume of $N_i$. Thus, the attracting
field of the core of $N_i$ is partially screened by quarks
belonging to neighboring nucleons. It
follows that quarks of $N_i$ "see"
a weaker attracting field and settle in a larger volume. As the
number of nucleons of a nucleus $A$ increases, the average
number of neighbors of a nucleon increases too and the screening
effect becomes more significant. This situation explains the
EMC effect.

\vglue 0.666666in
\noindent
{\bf 8. The Nuclear Tensor Force}
\vglue 0.333333in

   The equations of motion $(\!\!~\ref{eq:L2})$ and
$(\!\!~\ref{eq:L3})$ indicate that static electric field
of a charge and electric field of a moving monopole
have different dynamical properties. The same conclusion
holds for the corresponding magnetic fields. A special
case is found in the electric field of a polar dipole (which
is made of two displaced electric charges
having equal strength and opposite
sign) and that of an axial electric dipole of a spinning monopole.
The {\em axial} electric dipole of spinning monopoles is
discussed in this section.

    The neutron is known to be a spin-1/2 electrically neutral
composite particle. Its nonvanishing magnetic dipole moment
demonstrates that not all effects of its electrically charged
constituents vanish. The analogy between electric charges and
magnetic monopoles is the basis of the following statement.
If quarks are dyons and strong interactions are interactions
between magnetic monopoles then, by analogy with the
existence of the neutron's magnetic dipole moment
associated with electrically charged spinning quarks,
it is highly reasonable that
neutrons should have a large electric dipole moment
which is created by spinning monopoles. Indeed,
it is extremely unlikely that the overall electric
dipole moment of a system of spinning monopoles vanishes
whereas its total spin is nonzero. (It is evident that small
CP violations are irrelevant to the neutron's {\em axial}
electric dipole moment associated with spinning monopoles
which is completely analogous to the neutron's axial magnetic
dipole moment.) This discussion indicates that
the very low upper bound measured for the electric dipole
moment of the neutron[18,19] {\em looks} as a major argument against a
hadronic theory where quarks are magnetic monopoles.

    This argument {\em does not hold} in the case of
the charge-monopole
theory used here. This assertion relies on the following points.
As mentioned above in section 3, the
monopole theory derived in reference 5 concludes that charges
do not interact
with fields of monopoles. All experiments carried out for the
measurement of the electric dipole moment of the neutron are
eventually based on the interaction of an electric charge
(with which a static electric field is associated) with
the electric field of the searched electric
dipole moment of the neutron[18,19]. It should be pointed
out that according to (I) of section 3, charges interact
with {\em polar} electric dipoles associated with a distribution
of electric charges. On the other hand, they do not interact with
{\em axial} electric dipoles associated with spinning monopoles.
Thus, the very low upper bound measured for the electric dipole
moment of neutrons is, in fact, an upper bound for its {\em polar}
electric dipole moment. These measurements provide no information
on the magnitude of the neutron's {\em axial} electric dipole moment.
Hence, results on the measurements of the neutron's
electric dipole moment are not incompatible with the
monopole theory presented
in this work whose main results are $(\!\!~\ref{eq:L1})$,
$(\!\!~\ref{eq:L2})$ and $(\!\!~\ref{eq:L3})$.

    As pointed out above, a nucleon is expected to have a nonvanishing
axial electric dipole moment, due to its spinning quarks.
In this way, one finds an explanation for the tensor
interaction between nucleons[17]
\begin{equation}
V_T = \{3(\mbox{\boldmath $\sigma $}_1 \mbox{\boldmath $\cdot r$})
(\mbox{\boldmath $\sigma$}_2 \mbox{\boldmath $\cdot r$})
-r^2\mbox{\boldmath $\sigma $}_1\mbox{\boldmath $\cdot \sigma $}_2\}
U(r)
\label{eq:TENSOR}
\end{equation}
where {\boldmath $r$}={\boldmath $r$}$_2$-{\boldmath $r$}$_1$
and {\boldmath $\sigma $} is the spin operator. This
expression is a generalization of the dipole-dipole interaction
between two static point dipoles {\boldmath $\mu $}$_1$ and
{\boldmath $\mu $}$_2$[20]
\begin{equation}
V_{DIPOLE} = -\{3(\mbox{\boldmath $\mu $}_1
\mbox{\boldmath $\cdot r$})
(\mbox{\boldmath $\mu $}_2 \mbox{\boldmath $\cdot r$})
-r^2\mbox{\boldmath $\mu $}_1\mbox{\boldmath $\cdot \mu $}_2\}
/r^5
\label{eq:DIPDIP}
\end{equation}
Evidently, the nuclear tensor interaction
cannot be exactly a dipole-dipole one because nucleons are
not point dipoles but composite
particles whose size is not much smaller than the distance between
nucleons in a nucleus.
For this reason the form of the function $U(r)$ of
$(\!\!~\ref{eq:TENSOR})$ is determined phenomenologically.

    It is interesting to note that the {\em sign} of the
nuclear tensor force[17,21] is like that of two dipoles whose
moment bear the same relation to the particles' spin,
namely, both moments are either parallel or antiparallel
to the spin.
Moreover, the strength of the tensor interaction is greater
than that which exists between two magnetic dipoles associated
with spinning charges, like the nucleons magnetic dipoles.
This property is compatible with the behavior of spinning
monopoles that have an {\em axial} electric dipole moment.

\vglue 0.666666in
\noindent
{\bf 9. Exotic States of Hadrons}
\vglue 0.333333in

   The approach presented in this work provides a simple explanation
why $q\bar{q}$ and $qqq$ systems form tightly bound states.
In mesons, which
are bound states of a $q\bar {q}$ system, the
two quarks have the same absolute value of magnetic charge. Therefore,
due to the opposite signs of these charges, the corresponding
(magnetic) Coulomb
force holds them tightly together. The system as a whole is
neutral with respect to magnetic charges. The same
is true for baryons.
The magnetic monopole charge of three quarks, each of which has the
amount of $-g_0$,  balances the magnetic charge of the baryonic core
which is $3g_0$. Hence, quantum numbers of mesons correspond
to a pair of $q\bar {q}$ and those of baryons are created by $qqq$.
The existence of sea quarks comply with these conclusions because
all terms of a quantum state of a particle should have
the same quantum numbers.

   This structure of hadrons indicates that bound states of
two hadrons take place only due to {\em residual} forces,
like the nuclear ones. In most nuclei,
the interaction energy of these forces is about
8 Mev per nucleon.
In particular, there should be no {\em strongly} bound
state of a baryon and a meson, of two mesons, of two baryons,
of a baryon and an antibaryon
etc. These results are compatible with
experimental data.

\vglue 0.666666in
\noindent
{\bf 10. Concluding Remarks}
\vglue 0.333333in

    The cornerstone of the approach presented in this work is the
new classical theory of charges and monopoles[5] whose
basic results are given in the Lagrangian
$(\!\!~\ref{eq:L1})$ and the equations
of motion  $(\!\!~\ref{eq:L2})$
and $(\!\!~\ref{eq:L3})$. This theory is based on
some simple postulates, like the existence of a regular
Lagrangian density from which the equations of motion are derived. As
part of a classical theory,
these postulates are apparently irrelevant to
strong interactions. In other words, the main postulate
is based on general principles and is not
adopted in order to fit specific data obtained from
experiments with strongly interacting systems.
The consequences of this theory are examined here
and their compatibility with
basic experimental evidence is encouraging. The first
conclusion (I) says that there is no direct
interaction of charges with
fields of monopoles. The second conclusion shows that the elementary
unit of monopoles is a free parameter. The third one predicts
the existence of just two related interactions that conserve
flavor and parity.

    The first of these conclusions is
relevant to the well established
experimental fact that charged leptons do not participate in
strong interactions. As discussed in section 8, it
also removes a major objection based upon
the very low upper bound measured for the electric dipole moment
of the neutron. The second result pertains to the presently
accepted elementary magnetic charge $g_0^2=1/4e^2\simeq 34$.
A value like this looks
too high for hadronic calculations. Obviously, the usefulness
of $g_0$ as a {\em free} parameter is much better than
the huge {\em fixed}
value $g_0^2 \simeq 34$. The third prediction is satisfied
completely by Nature where one finds that only electromagnetic
and strong interactions satisfy flavor and parity conservation.

   The theory based on the Lagrangian
density $(\!\!~\ref{eq:L1})$
is of a very general character and additional postulates are needed
in order to fit specific properties of Nature.
These postulates represent basic properties of hadrons
and are used for the construction of the actual
model described in this work. They state
that the elementary unit of magnetic charge
is much greater than the electric one, that
baryons' quantum numbers are determined by configurations of
three valence quarks
and that the net magnetic charge of hadrons
vanishes.

    The baryonic core utilized in this work results from the
postulate stating that the overall magnetic charge of
hadrons vanishes.
This assumption paves the way for
the interpretation of the $qq$ force as a simple repulsive force
which is similar to the force between two electrons. In this case,
quarks, like electrons,  can be treated as ordinary Dirac particles
and strong interactions are related to electromagnetic ones
by the well known transformations $(\!\!~\ref{eq:DUALITY})$.
The kind of baryonic core used here also explains
the experimental data concerning the
portion of nucleonic linear momentum carried by quarks as well
as the relatively small Fermi motion of antiquarks in nucleons.

    Another conclusion found above is that electrons interact
differently with protons and neutrons whereas these two
kinds of nucleons are expected to be
alike in experiments with hard $\gamma $ photons. As shown in
the fifth section, this prediction is supported by experimental
data.

   Qualitative properties of nuclei are also explained
here. The similarity between nuclear forces and van der Waals
ones; the EMC effect where nucleons swell in nuclei;
the existence, sign and order of magnitude
of the nuclear tensor force are discussed.

    It is also found above that there should be no exotic states
of strongly bound hadrons. Thus, experiments should yield no
{\em strongly}
bound states of two mesons, of two baryons,
of a baryon and an antibaryon and of a
baryon and a meson. These predictions are compatible
with experimental data.

    If the approach presented in this work is found useful
then a common foundation for strong and electromagnetic
interactions becomes an element of the theory.
This point is supported by the discussion carried out
in section 2 which shows common physical properties of these
two kinds of interactions, namely flavor
and parity conservation. Another conclusion obtained is that
Maxwell's theory of electric and magnetic fields is extended
and these fields take a completely symmetric form, as shown in
the Lagrangian $(\!\!~\ref{eq:L1})$. The goal of showing a common
foundation for strong and electromagnetic interactions is one
of the motivations of Schwinger (see the second paper of reference 2)
and of Sawada (see most of the papers of reference 4) for their
work on dyon physics.

    As stated in the introduction,
the present work discusses only some
qualitative properties of hadrons.
Its main objective is to show that the
approach used here does not fail on this
elementary level of examination. Evidently, as a first
attempt, the scope of this work does not cover all qualitative
properties of hadronic systems.

\newpage
References:
\begin{itemize}
\item[{[1]}] P. A. M. Dirac, Phys. Rev. {\bf 74}, 817 (1948).
\item[{[2]}] J. Schwinger, Phys. Rev. {\bf 173}, 1536 (1968);
Science, {\bf 165}, 757 (1969).
\item[{[3]}] A. O. Barut, Phys. Rev. {\bf D3}, 1747 (1971).
\item[{[4]}] T. Sawada, Phys. Lett., {\bf b43}, 517 (1973);
Nuc. Phys., {\bf B71}, 82 (1974);
Phys. Lett., {\bf B52}, 67 (1974);
Prog. Theor. Phys., {\bf 59}, 149 (1978);
Prog. Theor. Phys., {\bf 63}, 2016 (1980);
Nuovo Cimento, {\bf A62}, 207 (1981);
Phys. Lett., {\bf B100}, 50 (1981);
Nuovo Cimento, {\bf A80}, 247 (1984).
\item[{[5]}] E. Comay, Nuovo Cimento, {\bf 80B}, 159 (1984).
\item[{[6]}] L. D. Landau and E. M. Lifshitz, The Classical Theory
of Fields (Pergamon, Oxford, 1975).
\item[{[7]}] H. J. Lipkin and M. Peshkin,
Phys. Lett. {\bf B179}, 109 (1986).
\item[{[8]}] E. Comay, Phys. Lett. {\bf B187}, 111 (1987).
\item[{[9]}] B. T. Feld, Models of Elementary Particles (Blaisdell,
Waltham Mass., 1969) p. 236.
\item[{[10]}] D. H. Perkins, Introduction to High Energy Physics
(Addison-Wesley, Menlo Park CA, 1987). pp 275-283.
\item[{[11]}] J. I. Friedman, Rev. Mod. Phys. {63} (1991) 615;
F. Halzen and A. D. Martin, Quarks and Leptons, An Introductory
Course in Modern Particle Physics (John Wiley, New York,1984)
p. 200.
\item[{[12]}] T. H. Bauer, R. D. Spital, D. R. Yennie and
F. M. Pipkin, Rev. Mod. Phys. {\bf 50} 261 (1978). (see pp. 269,293).
\item[{[13]}] R. D. Evans, The Atomic Nucleus (McGraw-Hill, New York,
1955) p. 365.
\item[{[14]}] W. D. Myers and W. J. Swiatecki, Ann. Phys. (NY),
{\bf 84}, 186 (1974).
\item[{[15]}] J. J. Aubert et al. (EMC), Phys. Lett. {\bf 123B},
275 (1983).
\newline
\hspace{0.4in}
A. Bodek et al., Phys. Rev. Lett. {\bf 50}, 1431 (1983).
\item[{[16]}] R. Sawafta et al., Phys. Lett {\bf 307B}, 293 (1993).
\item[{[17]}] A. deShalit and H. Feshbach, Theoretical Nuclear
Physics (John Wiley, New York, 1974). Vol 1, pp. 3-14.
\item[{[18]}] N. F. Ramsey, Ann. Rev. Nucl. Part. Sci.
{\bf 32}, 211 (1982);
Ann. Rev. Nucl. Part. Sci. {\bf 40}, 1 (1990).
\item[{[19]}] D. Dubbers in Progress in Particle and Nuclear
Physics, ed. A. Faessler (Pergamon, Oxford, 1991) p. 173.
\item[{[20]}] J. D. Jackson, Classical Electrodynamics (John
Wiley, New York, 1975). p. 143.
\item[{[21]}] J. M. Blatt and V. F. Weisskopf, Theoretical
Nuclear Physics (John Wiley, New York, 1952). p. 103.
\end{itemize}
\newpage
\noindent
Table Captions

\noindent
Table 1:

   Validity of flavor and parity conservation under three
kinds of interactions.
\newpage

\hspace{1.2in} Table 1.
\vglue 0.5in
\begin{tabular}{|c|c|c|c|} \hline
       & strong & electromagnetic & weak  \\ \hline
flavor & yes    & yes             & no    \\ \hline
parity & yes    & yes             & no    \\ \hline
\end{tabular}
\end{document}